\documentstyle {article}
\font\script=cmmib10%eusm10
\font\bbold=cmmib10%msbm10

\newcommand{\SC}{\mbox{\script C}}
\newcommand{\SS}{\mbox{\script S}}

\newcommand{\SM}{\mbox{\script M}}
\newcommand{\BR}{\mbox{\bbold R}}
\begin{document}
\textwidth 18.0cm
\textheight 23.0cm
\topmargin -0.5in
\baselineskip 16pt
\parskip 18pt
\parindent 30pt
\title{ \large \bf Geometry of chaos in the two-center problem in
General Relativity}
\author{Ulvi Yurtsever \\
~~~~~~~~~~~~ \\
Jet Propulsion Laboratory 169-327\\
California Institute of Technology\\
4800 Oak Grove Drive\\
Pasadena, CA 91109\\
and\\
Theoretical Astrophysics 130-33\\
California Institute of Technology\\
Pasadena, CA 91125}
\date{December, 1994\thanks{Submitted to Physical Review D}}
\baselineskip 24pt
\maketitle
\thispagestyle{empty}
\vspace{.2in}
\baselineskip 12pt
\begin{abstract}
\noindent The now-famous Majumdar-Papapetrou exact solution of the
Einstein-Maxwell equations
describes, in general,  $N$ static, maximally
charged black holes balanced under mutual gravitational and
electrostatic interaction.
When $N=2$, this solution defines the two-black-hole spacetime,
and the relativistic two-center problem is the problem
of geodesic motion on this static background.
Contopoulos and a number of other workers have recently discovered
through numerical experiments that
in contrast with the Newtonian two-center problem, where the dynamics
is completely integrable,
relativistic null-geodesic motion on the two
black-hole spacetime exhibits chaotic behavior. Here I identify
the geometric sources of this chaotic dynamics by first reducing the
problem to that of geodesic motion on a negatively curved (Riemannian)
surface.
\vspace{0.5cm}

\end{abstract}
\newpage
\baselineskip 14pt
\parskip 10pt
\pagestyle{plain}
\pagenumbering{arabic}
~~~~~~

{\noindent \bf 1. The Majumdar-Papapetrou solution}

The general Reissner-Nordstrom metric
\[
g \, = \, - \left( 1 - \frac{2M}{r} + \frac{Q^2}{r^2} \right)
\, dt^2 \, + \, {\left( 1 - \frac{2M}{r} + \frac{Q^2}{r^2} \right)}^{-1}
\, dr^2 \, + \, r^2 \left( d\theta^2 + \sin^2\theta \, d\phi^2 \right)
\]
for a charged black hole of mass $M$ and
charge $Q$ takes a particularly simple form when the black hole
is extremal with $|Q|=M$:
\begin{equation}
g \, = \, -\left( 1 - \frac{M}{r} \right)^2 \, dt^2 \, + \,
\left( 1 - \frac{M}{r} \right)^{-2} \, dr^2 \, + \,
r^2 \left( d\theta^2 + \sin^2\theta \, d\phi^2 \right) \;.
\end{equation}
In the isotropic coordinates $\bar{r} \equiv r-M$, Eq.\,(1) can be
written suggestively as
\begin{equation}
g \, = \, - \left( 1 + \frac{M}{\bar{r}} \right)^{-2} \, dt^2
\, + \, \left( 1 + \frac{M}{\bar{r}} \right)^2 \,^{(3)}\eta \; ,
\end{equation}
where
\[
\,^{(3)}\eta = d\bar{r}^2 + \bar{r}^2 \left( d\theta^2 + \sin^2 \theta
\,d\phi^2 \right)
\]
denotes the flat Euclidean metric. The metric function $1+M/\bar{r}$
appearing in Eq.\,(2) has the form of a harmonic function in Euclidean
space, and, miraculously, when $1+M/\bar{r}$ is replaced
with a more general harmonic function the metric Eq.\,(2)
still remains a solution to the Einstein-Maxwell equations ([1--2]).
More precisely, as first discovered by Majumdar and Papapetrou,
the metric
\begin{equation}
g \, = \, - U^{-2} \, dt^2 \, + \,
U^2 \left( dx^2 + dy^2 + dz^2 \right) 
\end{equation}
and the electromagnetic potential $A_a$ given by
\begin{equation}
A \, = \, \pm \frac{1}{U} \, dt 
\end{equation}
are a solution to the source-free Einstein-Maxwell equations
as long as the function $U=U(x,y,z)$
satisfies Laplace's equation in flat
space:
\begin{equation}
\sum_{k=1}^{3} U_{,kk} \, = \, U_{,xx}+U_{,yy}+U_{,zz} \, = \, 0 \; .
\end{equation}
Note that this solution is static ($\partial/\partial t$ is a timelike
Killing vector), but in general has no other symmetries.

It was first realized by Hartle and Hawking ([3]) that with the choice
\begin{equation}
U(\vec{r}) \, = \, 1 \, + \,
\sum_{i=1}^{N} \frac{M_i}{|\vec{r}-\vec{r}_i|}
\end{equation}
for the potential $U(x,y,z)$, the Majumdar-Papapetrou solution
represents $N$ extremal black holes, where the $i$'th
black hole, stationary at the fixed position
$\vec{r}=\vec{r}_i$, has mass $M_i$ and charge $|Q_i |=M_i$. All
charges $Q_i$ have the same sign given by the sign chosen in
Eq.\,(4), which ensures that the holes
remain in equilibrium, balanced under mutual
gravitational attraction and electrostatic repulsion.
The apparent singularity in $U(x,y,z)$
[and therefore in the metric Eq.\,(3)] at the positions
$\vec{r}=\vec{r}_i$ is the usual coordinate singularity associated with
static coordinates at an event horizon. Indeed, the surface area of a
small coordinate sphere $\{ t = {\rm const.} , |\vec{r}-
\vec{r}_i|=\epsilon \}$ around $\vec{r}=\vec{r}_i$ approaches the
expected surface area of the horizon:
\begin{eqnarray}
\lim_{\vec{r}\rightarrow \vec{r}_i} \left( U^2 4 \pi |\vec{r}-\vec{r}_i|^2
\right)
& = & 4 \pi \lim_{\vec{r}\rightarrow \vec{r}_i}
\left[ \left( 1+ \frac{M_i}{|\vec{r}- \vec{r}_i|} + {\rm O}(1)
\right)^2 |\vec{r}-\vec{r}_i|^2 \right] \nonumber \\
& = &
4 \pi {M_i}^2 \; ,
\end{eqnarray}
and the metric can be extended analytically into the interiors of the
black holes (into ``negative $|\vec{r}-\vec{r}_i|$") using Kruskal-like
coordinates. As in the single black hole case
(the extremal Reissner-Nordstrom solution), the interiors of the black
holes house true physical
singularities where spacetime curvature blows up.

~~~~~~

{\noindent \bf 2. Chaos in the two-black-hole spacetime}

When $N=2$, the spacetime given by Eqs.\,(6) and (3) represents a
relativistic analogue to the
two-center configuration in Newtonian gravity,
in which the Newtonian gravitational field is
generated by two point masses at fixed positions
(i.e., the mutual gravitational interaction of the masses is ignored).
Numerical investigations of null geodesic motion on this two-black-hole
spacetime by Contopoulos and coworkers ([4]) have revealed that
the geodesics exhibit chaotic behavior in the vicinity of the two
centers. More specifically, Contopoulos studies null geodesics
whose spatial motion is confined
to a two-dimensional symmetry plane; assuming the black holes
are positioned along the $z$-axis, this plane is
typically the surface $\{ x=0 \}$ (see Fig.\,1).
Numerical integration of the
null geodesic equations then reveals that for geodesics that
approach the black holes from infinity, it is
essentially impossible to predict whether the orbit will
plunge into the first
hole, or the second one, or escape back out to infinity; in other words
the qualitative behavior of the orbits near the black holes
exhibits effectively stochastic features.
This places the relativistic two-center motion
in surprising contrast with the corresponding Newtonian problem
(i.e., the motion of a massive test body
in the gravitational field of two
fixed centers) where the dynamics
is known to be completely integrable (a classical result that goes
back to the work of Jacobi and Liouville).

In this paper I will argue that the chaotic behavior
of the null geodesic flow has
its roots in the spatial geometry of the two-black-hole spacetime,
and I will do so by first showing that the dynamics
of this flow can be reduced to that of ordinary geodesics on a
negatively curved Riemannian surface.

~~~~~~
\newpage

{\noindent \bf 3. Geometric analysis of the two-black-hole
null geodesic flow}

I will rely on the well-known ``Fermat's principle" in
its relativistic formulation ([5]).
Fermat's principle states that if $\SM = \BR \times \Sigma$
is a static spacetime with metric
\[
g \, = \, g_{00}\, dt^2 \, + \,^{(3)}h \; ,
\]
where $\Sigma$ is a 3-manifold, and $g_{00} < 0$ is
a smooth function and $^{(3)}h$ is a Riemannian metric on $\Sigma$
(both independent of $t$), then the null geodesics of $(\SM ,g)$
when projected
onto $\Sigma$ are precisely the Riemannian geodesics of the 3-geometry
\begin{equation}
\left( \Sigma , \frac{^{(3)}h}{-g_{00}} \right ) \; ;
\end{equation}
and, furthermore, the affine parameter (i.e., the arc
length) along the projected geodesics in $[\Sigma , \,^{(3)}h/(-g_{00})]$
is precisely the static time coordinate $t$ measured
along the null geodesics in $(\SM ,g)$. In words that would have sounded
familiar to Fermat, the principle states that light follows the path of
shortest (or extremal) travel time between two given points in 3-space.

In the multi-black-hole solution given by Eqs.\,(3) and (6), Fermat's
principle shows that null geodesic flow in the asymptotically
flat exterior region (outside the event horizons of the
black holes) is equivalent to the Riemannian geodesic flow
of the 3-geometry $(\Sigma ,h)$,
with the 3-manifold $\Sigma$ given by
\mbox{$\Sigma = \BR^3 \setminus \{ N \; {\rm points} \} $},
and with the Riemannian metric $h$ on $\Sigma$ given by
\begin{equation}
h \, \equiv \, \frac{^{(3)}h}{-g_{00}} \, = \,
\Omega^2 (dx^2 +dy^2 +dz^2 ) \; ,
\end{equation}
where
\begin{equation}
\Omega \, = \, U^2 \, = \, \left( 1 + 
\sum_{i=1}^{N} \frac{M_i}{|\vec{r}-\vec{r}_i|} \right)^2 \; .
\end{equation}
In the two-black-hole spacetime, I can assume without loss of
generality that the holes are positioned along
the $z$-axis at $\vec{r}_1 = (0,0,1)$ and
$\vec{r}_2 = (0,0,-1)$ (see Fig.\,1). It is obvious that any two-plane
containing the symmetry ($z$-) axis is a totally geodesic submanifold of
$\Sigma$. As I will focus on null geodesics which lie (spatially)
in such a symmetry plane, which I can assume to be
the $yz$-plane $\{ x=0 \}$ as in Fig.\,1, by Fermat's principle
the null geodesic flow I need to study is equivalent to the
geodesic flow on the two-dimensional Riemannian surface $(\SS ,h)$,
where \mbox{$\SS = \BR^2 \setminus \{ (0,1),(0,-1) \} $}, and
\begin{eqnarray}
h & = & \Omega^2 (dy^2 + dz^2 ) \; ,  \\
\Omega & = & \left( 1 + \frac{M_1}{\sqrt{y^2 + (z-1)^2}}
+ \frac{M_2}{\sqrt{y^2 + (z+1)^2}} \right)^2 \; .
\end{eqnarray}
Note that the geodesic flow of $(\SS ,h)$ corresponds,
in the original spacetime, only to the null geodesic
flow in the exterior of the black holes; null geodesic motion in the
interior regions is not covered by this correspondence. This point
will become clearer after a closer
look at the topology and large-scale geometry of $(\SS ,h)$:

{\noindent \bf Global geometry of the Riemannian surface $(\SS ,h)$}

Look closely at the behavior of the metric $h$ near the centers, e.g.,
near $\vec{r}=\vec{r}_1=(0,1)$. Introducing Euclidean polar coordinates
$(R, \theta )$ centered around $\vec{r}_1=(0,1)$ (i.e., $R \equiv
|\vec{r}-\vec{r}_1 |$), I can write the conformal factor $\Omega$ in the
vicinity of $\vec{r}_1$ as
\begin{equation}
\Omega \, = \, \left( 1 + \frac{M_2}{2} + \frac{M_1}{R} + O(R) \right)^2
\; ,
\end{equation}
and similarly I can write
\begin{equation}
h \, = \, \left( 1 + \frac{M_2}{2} + \frac{M_1}{R} + O(R) \right)^4
(dR^2 + R^2 \, d \theta^2 ) \; .
\end{equation}
Now introduce a new radial coordinate $\rho \equiv {M_1}^2 /R$. Then
Eq.\,(14) becomes
\begin{eqnarray}
h & = & \left( 1 + \frac{M_2}{2} + \frac{\rho}{M_1}
+ O(\frac{1}{\rho} ) \right)^4 \frac{{M_1}^4}{\rho^4}
(d \rho^2 + \rho^2 \, d \theta^2 ) \nonumber \\
& = & \left( 1 + O(\frac{1}{\rho}) \right)^4
(d \rho^2 + \rho^2 \, d \theta^2 ) \; .
\end{eqnarray}
A similar analysis can be carried out in the vicinity of the other
center $\vec{r}=\vec{r}_2$ with the same conclusion, namely that
what looks like a singularity at $\vec{r}=\vec{r}_1$
(and similarly near the other center) is in fact an entire
asymptotically flat Euclidean region squeezed into a small neighborhood
of the ``point" $\vec{r}_1=(0,1)$ in the coordinate system $(y,z)$.
The global geometry of the surface $(\SS ,h)$ is then as
depicted in Fig.\,2 below, with three asymptotically flat regions, one
at $\vec{r} \rightarrow \infty$, and two others
at each of the centers $\vec{r} \rightarrow \vec{r}_1$ and
$\vec{r} \rightarrow \vec{r}_2$. As a
corollary, the surface $(\SS ,h)$ is geodesically
complete. This is expected, since
by Fermat's principle the affine parameter (i.e., arc length) along the
geodesics of $(\SS ,h)$ is the static time coordinate $t$
measured along the null
geodesics of the two-black-hole spacetime, and static time diverges to
infinity at the event horizons of the black holes. In other words,
a null geodesic in the two-black-hole spacetime falls into the $i$'th
black hole if and only if
the corresponding Riemannian geodesic in $(\SS ,h)$
escapes into the asymptotic region $\vec{r} \rightarrow \vec{r}_i$.

{\noindent \bf Local geometry of the Riemannian surface $(\SS ,h)$}

The intrinsic geometry of a two-dimensional Riemannian manifold
{\sl in the small} is determined completely by the Gaussian
curvature $K$ (which is one-half the scalar curvature $R$).
With the metric written in the conformally flat form
Eq.\,(11), $K$ is given by
\begin{eqnarray}
K & = & - \frac{1}{\Omega^2} \, \triangle (\log \Omega ) \\
& = & \frac{1}{\Omega^4} \left[ \Omega^3 \left(
(\Omega^{-1})_{,yy} + (\Omega^{-1})_{,zz} \right) -{\Omega_{,y}}^2
-{\Omega_{,z}}^2 \right] \; ,
\end{eqnarray}
where $\triangle$ denotes the scalar Laplacian in the flat metric $dy^2
+ dz^2$. It is straightforward to compute $K$ for the surface $(\SS ,h)$
by simply substituting $\Omega$ from Eq.\,(12) in Eq.\,(17). The result
is a complicated expression, not particularly illuminating in its analytic
form (which therefore I will not bother to give).
A plot of the curvature $K$ as a function of
the coordinates $y, \; z$ is given in Fig.\,3
(where I chose unit masses $M_1 = M_2 =1$). It is apparent that $K$ is
strictly negative throughout $\SS$ (and this is true for all masses $M_1
, \; M_2 > 0$). Both far away from and near the centers
(where geometry is asymptotically flat)
$K$ approaches zero from below as expected (see Fig.\,2).

~~~~~~

{\noindent \bf 4. Can chaos in the two-black-hole spacetime
be explained solely by the negatively curved geometry of $(\SS ,h)$?}

In a Riemannian manifold of
arbitrary dimension $n$, negative sectional curvature causes neighboring
geodesics to diverge exponentially ([6,7]).
Recall the derivation of this well-known result: If $Z$ denotes a vector
field along the geodesic $\gamma$, Lie transported by a congruence of
neighboring geodesics, then 
\begin{equation}
{\nabla}_{\gamma_{\ast}} {\nabla}_{\gamma_{\ast}} Z
\, = \, {R_{\gamma_{\ast}}}_{Z} \, \gamma_{\ast} \; ,
\end{equation}
where $R_{XY}$ denotes the curvature operator ${\nabla}_X \nabla_Y
-\nabla_Y \nabla_X - \nabla_{[X,Y]}$. Along $\gamma$,
an infinitesimal neighboring
geodesic can then be defined abstractly as any solution of
the ``Jacobi equation;" a differential equation along
$\gamma$ derived from Eq.\,(18). In a
parallel-propagated basis $\{ E_k \}$ along $\gamma$ such that $E_n
=\gamma_{\ast}$, Jacobi's equation is
\begin{equation}
\frac{d^2 Z^a}{ds^2} \, = \, - {R^a}_{nbn} Z^b \; ,
\end{equation}
where $s$ is the affine parameter. For a two-dimensional Riemannian
manifold with Gaussian ($\equiv$ sectional) curvature $K=R_{1212}$,
and with $Z \equiv Z^1$,  Eq.\,(19)
becomes
\begin{equation}
\frac{d^2 Z}{ds^2} \, = \,- K \, Z \; . 
\end{equation}
Assuming $K<0$, and assuming the affine parameter
$s$ is small compared to $|K/(dK/ds)|_{s=0}|$,
Eq.\,(20) has generic solutions of the form
\begin{equation}
Z(s) \, \sim \, A(s) \exp \left( \int^{s} \sqrt{-K} \, ds' \right) \, + \,
B(s) \exp \left( - \int^{s} \sqrt{-K} \, ds' \right) \; ,
\end{equation}
where $A(s)$ and $B(s)$ are slowly-varying amplitudes. It is clear that
negative Gaussian curvature
$K$ results in an exponentially diverging $Z(s)$ in general. When
$K(s)$ is bounded from above by a negative number,
Eqs.\,(20)--(21) imply that
$\gamma$ (as an orbit in the geodesic flow) has positive Liapunov
exponents. Obviously, exponential instability of orbits and positive
Liapunov exponents are sufficient conditions for the presence of
``sensitive dependence on initial conditions," the key ingredient of
chaos. But are these criteria sufficient to demonstrate that chaotic
behavior is indeed present?

To investigate this question, let me briefly consider two examples from
Newtonian gravitation. Recall that in classical mechanics,
for a Hamiltonian system with Lagrangian function
\[
L \, = \, \mbox{$1 \over 2$} \sum_{j,k}
a_{jk} {\dot{q}}^j {\dot{q}}^k \, - \, V(q^i) \; ,
\]
motion on a constant-energy ($\equiv$ constant-Hamiltonian) surface $\{
H=E \}$ is equivalent to geodesic motion on a Riemannian
manifold, namely on the submanifold
$\{ V(q^i ) < E \}$ of configuration space equipped
with the Riemannian metric
\begin{equation}
g_E \, = \, \left[ E - V(q^i ) \right]
\sum_{j,k} a_{jk} \, dq^j \otimes dq^k 
\end{equation}
(Hamilton-Jacobi-Maupertius--$\cdots$ principle; see [6]). Accordingly,
in mathematical analogy with relativistic gravitation, so also
in Newtonian gravity test-particle dynamics
has a geometric description in terms of geodesic motion.
In particular, motion in the Kepler and
Newtonian two-center problems can
both be described in terms of geodesics on a
two-dimensional Riemannian surface, and
this description can be put in
exactly the same form as in Eqs.\,(11)--(12), except, of course, in the
Kepler case the conformal factor $\Omega$ takes the form
\begin{equation}
\Omega \, = \, \left( E + \frac{M}{r} \right)^{1/2} \; ,
\end{equation}
and in the Newtonian two-center problem it has the form
\begin{equation}
\Omega \, = \, \left( E + \frac{M_1}{|\vec{r}-\vec{r}_1 |}
+ \frac{M_2}{|\vec{r}-\vec{r}_2 |} \right)^{1/2} \; .
\end{equation}
Plots of the Gaussian curvature of the metric
$g_E$ [Eq.\,(22)] for the (planar) Kepler and
Newtonian two-center problems are shown in Fig.\,4. In both plots, $E$
is chosen to be $E=-0.1$ ($E$ is chosen
negative so that the geodesic flow describes
the motion of bound orbits), and the masses are $M=M_1 = M_2 = 1$.
There are no surprises: As both systems are completely integrable with
stable closed orbits, one would not expect negative curvature to be
the dominant geometric feature. Indeed, in the Kepler case curvature is
strictly positive, and in the two-center case it is mostly
positive, with a small neighborhood
of negative curvature in the vicinity of
the centers; this small region of negative $K$
corresponds to directional instabilities the orbits
have while passing in between the two centers of attraction.
[Note that in contrast
with the black-hole surface, the center(s) in the Newtonian case are
genuine singularities of the metric $g_E$; however, these are not
curvature singularities ($K$ remains bounded as $\vec{r} \rightarrow
\vec{r}_i$), but rather conical singularities
with a mass-independent angle deficit $\pi$.]

So far the association between negative Gaussian curvature
on the one hand and chaotic behavior
of the geodesic flow on the other
appears to hold within the context of the
three examples I discussed. Consider, however, one more example, this
time the geodesic flow on the Riemannian surface $\SS$ with only
one extremal (Reissner-Nordstrom) black-hole; in other words with
metric $h$ given by Eq.\,(11) where $\Omega = (1+M/r)^2$. The Gaussian
curvature $K$
of the resulting geometry is plotted in Fig.\,5 (with
$M=1$). As in the
two-black-hole case (Fig.\,3), $K$ is strictly negative everywhere.
But the geodesic flow on this surface is a completely integrable system
(angular momentum provides the second integral of motion). Clearly,
then, negative curvature (sensitive dependence on initial conditions) is
{\sl not} sufficient for chaos: In fact, the unique closed (unstable)
geodesic in the geometry of Fig.\,5 has strictly positive Liapunov
exponents as an orbit in the flow, so even the
presence of positive Liapunov exponents
does not always imply chaotic behavior.

As others have done before, I
would like to argue in this paper against the widespread practice in the
physics literature of identifying chaos with merely the presence of
positive Liapunov exponents. This is especially important in relativity
(where there is no canonical choice for dynamical time) since whether
or not a Liapunov exponent is positive depends crucially on the nature of
the time parameter used in defining the exponent. In the next section I
will describe a precise formulation for ``chaos" [due to S.\ Willard
([8])] which I believe is particularly useful in relativity since it
does not depend sensitively on the choice of time. In the following
section (Sect.\,6), I will demonstrate that
null geodesic flow in the two-black-hole
spacetime is chaotic according to this formulation.

~~~~~~

{\noindent \bf 5. A precise formulation of chaos}

Central to our intuitive understanding of chaotic behavior is the notion
of ``sensitive dependence on initial conditions:" long-time prediction of
motion in the phase space of a chaotic system is impossible since small
initial perturbations of the orbits grow arbitrarily large as the
system evolves in time. This, of course, is a vague idea in need of
a precise mathematical formulation, and there exist various such
formulations, the concept of Liapunov exponents being one of them.
However, the
exact content of our intuitive notion of sensitive dependence is not
fully captured by the more precise concept of positive
Liapunov exponents. For
example, the phase flow $\{ \dot{x}=x , \; x \in \BR^n \}$ has
positive Liapunov exponents along all its orbits, but, clearly, this
is not a chaotic system, and more complicated ``counterexamples"
with positive exponents can be found in which to discern that motion
is non-chaotic would not be so easy.
In order to conclude, on the basis of the presence of
positive Liapunov exponents, that chaos is present,
it is apparently necessary to make sure that the
divergence of nearby orbits does not
occur simply because these orbits escape to ``infinity"
under time evolution. What is needed to address this
point is a mathematical formulation
slightly more sophisticated than the concept of
Liapunov exponents.

Here, then, is my favorite ``definition" of chaos, adopted from [8]:
Restrict attention, for definiteness, to phase spaces $\SM$ with
metrizable topology.
A dynamical system $(\SM , \varphi_t )$ is chaotic if it contains
a ``chaotic invariant subset," that is, a subset $\Lambda
\subset \SM$ \vspace{5pt}such that:
\newline (C1) $\Lambda$ is compact, and invariant under $\varphi_t$,
i.e., \vspace{5pt}$\varphi_t (\Lambda ) \subset \Lambda \; \;
\forall t \in \BR$.
\newline (C2) $\Lambda$ has sensitive dependence on initial
\vspace{5pt}conditions.
\newline (C3) $\Lambda$ is topologically \vspace{5pt}transitive.
\newline The precise meaning of condition C2 (sensitive dependence
on initial conditions) is the following: Fix a
distance function $\rho$ on $\SM$
compatible with $\SM$'s topology.
Condition C2 holds if there exists a {\sl fixed}
$\delta > 0$ such that for all $x \in \Lambda$ and for every
neighborhood $U\subset \Lambda$ of $x$
open in $\Lambda$, a point $y \in U$
and a $t>0$ can be found such that
\[
\rho [ \varphi_t (x) , \varphi_t (y) ] > \delta \; .
\]
In other words, given any point $x \in \Lambda$, no matter how small
a neighborhood $U$ of $x$ I choose I can always
find points $y \in U \cap \Lambda$
whose orbits eventually diverge away from that of $x$ under the
flow $\varphi_t$.
Since $\Lambda$
is compact, this notion of sensitive dependence on initial conditions
is independent of the choice of $\rho$. Topological transitivity
of $\Lambda$ (condition C3) means the following:
for every open $U, \; V \subset \Lambda$ there
exists a $t \in \BR$ such that \mbox{$\varphi_t (U)
\cap V \neq \emptyset $}.

Because the problem I study in this paper involves chaos ``localized" in
a bounded region of an asymptotically flat geometry (i.e., in the
vicinity of the black holes), I will need to use a slightly generalized
version of the above definition; my generalization
is designed to be adapted to the
essentially time-asymmetric nature of the problem (i.e., null
geodesics approaching the black-hole region from infinity and plunging
into the holes after exhibiting chaotic behavior). Namely, call
a subset $\Lambda
\subset \SM$ a ``chaotic future-invariant set" \vspace{5pt}if
\newline (FC1) $\Lambda$ is compact, and future-invariant
under $\varphi_t$,
i.e., \vspace{5pt}$\varphi_t (\Lambda ) \subset \Lambda \; \;
\forall t > 0$.
\newline (FC2) $\Lambda$ has sensitive dependence on initial
\vspace{5pt}conditions (defined as before).
\newline (FC3) $\Lambda$ is topologically \vspace{5pt}future-transitive.
\newline Note that topological transitivity of
$\Lambda$ as defined above (condition C3)
ensures essentially that the flow is topologically ``mixing;"
this condition is designed to rule
out situations in which $\Lambda$ can be decomposed into multiple
compact invariant sets. Clearly, topological
transitivity would be an inappropriately strong condition
to impose on a subset which is only
future-invariant. Therefore, I modify this condition so as to
demand that the flow on $\Lambda$ is
mixing only in the future direction, more precisely,
I define $\Lambda$ to be topologically future-transitive
if there exists a time $T>0$ such that for every pair of open subsets $U, \;
V \subset \varphi_T(\Lambda )$ times $t>0$ and $s>0$ can be found such that
$\varphi_t (U) \cap \varphi_s (V) \neq \emptyset$.
Clearly, a chaotic invariant set is also trivially a chaotic
future-invariant set.
The definition of a chaotic system can now be generalized to
include any dynamical system which contains
a chaotic future-invariant subset.
 
Notice that this definition for chaos makes no reference to Liapunov
exponents; in fact, the rate of divergence of nearby orbits is not
constrained in any way by the precise notion of sensitive dependence on
initial conditions. This fact makes the definition especially
interesting for applications in General Relativity: sensitive
dependence as defined above
holds for one choice of time function if and only if it holds
for any other, as long as two choices of time are always related by
a monotone-increasing diffeomorphism
from the real axis $\BR$ onto $\BR$. Of course, in
general a mathematical definition is
useful only if it is the subject of theorems, and there
do exist theorems which
demonstrate that many of the usual properties of chaotic systems can be
derived from the above conditions C1--C3 (or FC1--FC3); I will not
discuss these results here, but direct the reader to the literature,
especially as listed in [8]. Instead I
will turn now to the demonstration that
the geodesic flow on the two-surface $(\SS ,h)$ (which, as
I discussed in Sect.\,3, is equivalent to the null geodesic flow of the
two-black-hole spacetime)
is chaotic according to the formulation of chaos I just described.
It is important to note here that
other studies (see [9]--[10] and references therein)
have carried out this demonstration by searching for various more
direct signatures of chaos in the two-black-hole
geodesic flow; for instance, the
existence of hyperbolic cycles and transverse homoclinic
orbits in this flow is discussed in [9], and the
presence of positive Liapunov exponents is explored in [10].

~~~~~~

{\noindent \bf 6. ``Proof" of chaos in the two-black-hole null geodesic
flow}

The geodesic flow on the Riemannian surface $(\SS ,h)$ can be
described as a Hamiltonian dynamical system, with phase space $\SM$
$=$ the unit cotangent bundle of $\SS$, i.e.,
\begin{equation}
\SM \, = \, {T_1}^{\ast} \SS \, \equiv \, \{ (x,p) \in T^{\ast} \SS
\; | \; \Vert p \Vert =h^{ab} p_a p_b =1 \} \; ,
\end{equation}
and with the Hamiltonian function $H(x,p)= {1 \over 2} h^{ab}p_a p_b$.
I will denote the geodesic flow on $T_1^\ast \SS$ by the usual symbol
$\varphi_t$.
I showed in Sect.\,3 that $(\SS ,h)$ has strictly negative Gaussian
curvature, and recalled in Sect.\,4 that negative curvature
causes exponential divergence of the orbits in the geodesic flow.
Now, if the
surface $\SS$ were compact, I could then simply define
my invariant set $\Lambda$ to be
the entire phase space $\SM = T_1^\ast \SS \,$:
$\,$so chosen, $\Lambda$ is compact
when $\SS$ is, and because of the negatively curved geometry of $(\SS ,h)$,
$\Lambda$ has sensitive dependence on initial conditions,
i.e., satisfies condition C2 as formulated in the
previous section. It is not difficult to show also that $\Lambda$
is topologically transitive under the geodesic flow; therefore, if $\SS$
were compact, all conditions \mbox{C1--C3} for a chaotic invariant subset
would be satisfied by this simple
choice of $\Lambda$, i.e., the entire phase space would be a chaotic
invariant set. Indeed, it is well known that geodesic flows of
compact manifolds with negative
sectional curvature are chaotic.
(These flows in fact satisfy every criteria ever invented for chaos:
they have positive Liapunov exponents, positive entropy, are mixing,
are K-flows, $\cdots \;$ See [7] for an extensive but readable analysis
of this classical problem.)
The noncompactness of the two-black-hole Riemannian
surface $(\SS ,h)$ is then the main difficulty I
need to overcome in demonstrating
the existence of a chaotic (future-) invariant subset
in the (noncompact) phase space $T_1^\ast \SS$.

I will now construct a closed subset $\Lambda \subset T_1^\ast \SS$
which I claim is a chaotic future-invariant set for the geodesic flow.
That $\Lambda$ is compact and future-invariant will be evident from its
construction, however, I will
not be able to prove that $\Lambda$ satisfies conditions FC2 and
FC3. To prove these conditions, it would be sufficient to combine the
negatively curved geometry of $(\SS ,h)$ with the
intricate topological structure that $\Lambda$
appears to have; however, I cannot
prove that $\Lambda$ indeed has this intricate structure. As is usually
the case with studies of chaotic behavior, the evidence for this structure is
exclusively numerical. Some of this numerical
evidence I will present here,
and more of it can be found in the literature, e.g., in [4] and [10].

First define subsets $\Gamma_1, \; \Gamma_2$ and $\Gamma$ of the phase space
$T_1^\ast \SS$ as follows:
$\Gamma_i$ is the set of all points in $T_1^\ast \SS$ which fall into the
$i$'th black hole as $t \rightarrow \infty$, i.e.,
\begin{equation}
\Gamma_i \, \equiv \, \{ m \in T_1^\ast \SS \; | \;
\vec{r} \, [\varphi_t (m)] \longrightarrow \vec{r}_i \; {\rm as}
\; t \rightarrow \infty \} \; ,
\end{equation}
and $\Gamma$ is the set of all points which escape to the asymptotically
flat region $\vec{r}=\infty$ as $t \rightarrow \infty$, i.e.,
\begin{equation}
\Gamma \, \equiv \, \{ m \in T_1^\ast \SS \; | \;
\vec{r} \, [\varphi_t (m)] \longrightarrow \infty \; {\rm as}
\; t \rightarrow \infty \} \; . 
\end{equation}
Since these subsets consist of points $(x,p)$ such that the geodesic
starting at $x$
with initial tangent vector $p$ eventually escapes to one of the three
asymptotically flat regions of $(\SS ,h)$ (see Fig.\,2), it is clear
that both the $\Gamma_i$ and $\Gamma$ are open subsets in $T_1^\ast \SS$.
Also (and this will be important below), it is clear that $\Gamma$,
$\Gamma_1$ and $\Gamma_2$ are mutually disjoint subsets, i.e.,
\[
\Gamma \cap \Gamma_i \, = \, \Gamma_1 \cap \Gamma_2 \, = \, \emptyset \; .
\]
Now define $\Delta$ as the closed subset
\begin{equation}
\Delta \, \equiv \, {\rm complement} \,
(\Gamma \cup \Gamma_1 \cup \Gamma_2 ) \, = \,
(\Gamma \cup \Gamma_1 \cup \Gamma_2 )' \; ;
\end{equation}
$\Delta$ is the set of all points which
do not escape to any asymptotic region
as $t \rightarrow \infty$, i.e., the set of all future-imprisoned (e.g.,
periodic or quasi-periodic) orbits of the geodesic flow. This is
obviously a future-invariant subset, but it is not necessarily compact
(unless all imprisoned orbits are closed geodesics, which
is not the case as numerical studies show).
To cut $\Delta$ down to a compact size, introduce a compact subset $D
\subset \SS$ as follows (see Fig.\,6): Draw a circle
$\SC$ in the asymptotically flat region
$\vec{r} \rightarrow \infty$ which encloses both black holes,
and draw circles $\SC_1$ and $\SC_2$  in the
asymptotic regions $\vec{r} \rightarrow
\vec{r}_1$ and $\vec{r} \rightarrow
\vec{r}_2$ which enclose the black holes $1$ and $2$, respectively.
Choose these circles large enough so that if $\vec{n}$ denotes the
outward normal to $\SC$ and $\SC_i$, a geodesic $\gamma$
which crosses any one of the circles in the outward direction [i.e.,
with $h(\gamma_\ast , \vec{n} ) \geq 0$] escapes to the corresponding
infinity (and thus never crosses $\SC$ or $\SC_i$ again). By asymptotic
flatness, it is clear that such circles
$\SC$ and $\SC_i$ can be found (see Fig.\,6).
Now let $D \subset \SS$ be the compact
region bounded by the circles, in other words, define $D$ to
be the unique connected component of $\SS \setminus (\SC \cup
\SC_1 \cup \SC_2 )$ such that $\partial D = \SC
\cup \SC_1 \cup \SC_2$. Then put
\begin{equation}
\Lambda \, \equiv \, \Delta \cap T_1^\ast D \; .
\end{equation}
So constructed, $\Lambda$ is clearly both compact (a closed subset of a
compact set) and future-invariant.
I claim that this $\Lambda \subset T_1^\ast \SS$
is a chaotic future-invariant subset for the geodesic flow on
$T_1^\ast \SS$.

As I mentioned above, I am not able to prove that $\Lambda$ satisfies
the conditions FC2 and FC3 of Sect.\,5. Nevertheless,
a great deal of insight into the
structure of $\Lambda$ can be obtained by numerically
integrating the geodesic equations on $(\SS ,h)$. Extensive numerical
studies of this kind have been reported in [4] and [10].
Although I will base the
following observations on my own minimal investigation of the (numerical)
structure of $\Lambda$, these observations are supported by
the more extensive numerical evidence already published
in the literature.

Because of the exponential instability of all orbits in the geodesic
flow, it is clear that a direct computer proof of the existence of a
future-imprisoned orbit (lying in $\Lambda$)
is impossible: any real orbit in
the computer will eventually diverge away from $\Lambda$ because of
numerical instabilities, even if initially it lies in $\Lambda$. So it
might appear at first that by relying on numerical
integration it is impossible
to even prove that $\Lambda$ is nonempty! This is not the case, however;
numerical integration does yield an indirect proof
that orbits which lie in $\Lambda$ exist. More precisely, consider
those orbits whose starting points are on the $z$-axis and whose initial
(unit) tangent vectors are entirely
in the $y$-direction. [See Fig.\,7; all
orbits plotted in Fig.\,7 are of this kind. Also, although the orbits
plotted in Fig.\,7 are (mostly) with unit masses $M_1 =M_2 =1$ and
(some) with masses $M_1 =2$, $M_2 = 1$, similar behavior is observed
with all positive choices of $M_1 , \; M_2$.] $\,$In the
following, I will not make any distinction between points on the $z$-axis
and initial conditions for the orbits in $T_1^\ast \SS$; the
initial-tangent-vector part of the initial conditions is fixed
throughout to be a unit vector in
the $y$ direction. Now, by numerically
integrating these orbits into the future, the following features can
be observed: (i) Consider
any open interval of initial conditions (starting
points) on the $z$-axis lying in the vicinity of the
centers. No matter how small this interval is,
there are always points in it which belong to $\Gamma$,
$\Gamma_1$ and $\Gamma_2$. (ii) In any such interval, between any
two points that belong to a distinct pair of
the subsets $\Gamma$, $\Gamma_1$ and
$\Gamma_2$, there exists a third point
which belongs to the subset other than the two in the pair.

Note that since $\Gamma$, $\Gamma_1$ and $\Gamma_2$
are open sets, both of the statements
(i) and (ii) are ``stable" numerically, i.e., they can be verified with
arbitrarily-high-accuracy numerical calculations. Already, the statement
(i), combined with the observation that $\Gamma$ and $\Gamma_i$ are
disjoint, proves that $\Lambda$ is nonempty: a connected open interval in
$\BR$ cannot be the union of three disjoint open subsets, therefore, in
any interval of the kind described in (i) there must exist points which
belong to $\Lambda$. As I
remarked above, to prove that $\Lambda$ satisfies the conditions
FC2 and FC3 of Sect.\,5, it is sufficient to combine the exponential
instability of the geodesic flow on $T_1^\ast \SS$ with the
everywhere-dense topological structure of $\Lambda$, i.e., the structure
of a Cantor set of periodic or quasi-periodic orbits, so that every
open neighborhood of any point $m \in \Lambda$ contains points of
$\Lambda$ other than $m$. That $\Lambda$ indeed has this structure is
strongly suggested by the numerical evidence discussed here and more
extensively in [4] and [10]. However, the discovery of an analytical
proof of this topological structure remains an open problem.

{\bf \noindent Acknowledgements}

This research was carried out at the Jet
Propulsion Laboratory, Caltech, and was
sponsored by the NASA Relativity Office and by the National Research
Council through an agreement with the National Aeronautics and Space
Administration.

\newpage

\begin{center}
{\bf REFERENCES}
\end{center}

\noindent{\bf 1.} S.\ D.\ Majumdar, Phys.\ Rev.\ {\bf 72}, 390 (1947).

\noindent{\bf 2.} A.\ Papapetrou, Proc.\ R.\ Irish Acad.\
{\bf A51}, 191 (1947).

\noindent{\bf 3.} J.\ B.\ Hartle and S.\ W.\ Hawking, Commun.\ Math.\
Phys.\ {\bf 26}, 87 (1972).

\noindent{\bf 4.} G.\ Contopoulos, Proc.\ R.\ Soc.\ London {\bf A431},
183 (1990); {\bf 435}, 551 (1991). See also S.\ Chandrasekhar,
Proc.\ R.\ Soc.\ London {\bf A421}, 227 (1989).

\noindent{\bf 5.} C.\ W.\ Misner, K.\ S.\ Thorne and J.\ A.\
Wheeler, {\it Gravitation} (Freeman, San Francisco 1973).

\noindent{\bf 6.} V.\ I.\ Arnold, {\it Mathematical Methods of Classical
Mechanics} (Springer-Verlag, New York 1978).

\noindent{\bf 7.} V.\ I.\ Arnold and A.\ Avez, {\it Ergodic Problems of
Classical Mechanics} (Addison-Wesley, Redwood City 1989).

\noindent{\bf 8.} S.\ Willard, {\it Introduction to Applied Nonlinear
Dynamical Systems and Chaos} (Springer-Verlag, New York 1990).

\noindent{\bf 9.} Y.\ Sota, S.\ Suzuki and K.\ Maeda, Waseda University
Preprint, 1994.

\noindent{\bf 10.} C.\ P.\ Dettmann and N.\ E.\ Frankel, Phys.\ Rev.\ D
{\bf 50}, R618 (1994).

\newpage

\begin{center}
{\bf FIGURE CAPTIONS}
\end{center}

\noindent{\bf Figure 1.} Contopoulos's
(and also this paper's) analysis of the two-black-hole
null-geodesic flow is confined to those
null geodesics which lie in a two-dimensional
surface of symmetry such as the $yz$-plane $\{ x=0 \}$.

\noindent{\bf Figure 2.} The geometry of the Riemannian manifold
$(\SS ,h)$ in the large. Note that this is {\sl not} the actual
geometry of the surface $\{ x=0 \}$ in the physical metric $g$ on the
two-black-hole spacetime, but, rather, it is
the physical geometry with an extra conformal factor
introduced in accordance with Fermat's principle. In particular, only
the asymptotic region $\vec{r} \rightarrow \infty$ corresponds to the
usual asymptotic region in the physical spacetime; the asymptotic regions
$\vec{r} \rightarrow \vec{r}_i$ exist because of the singular behavior
of the static time coordinate $t$ at the event horizons of the black
holes. Accordingly, a null geodesic in the
two-black-hole spacetime falls
into the $i$'th black hole if and only if
the corresponding Riemannian geodesic in $(\SS ,h)$
escapes into the asymptotic region $\vec{r} \rightarrow \vec{r}_i$.

\noindent{\bf Figure 3.} The Gaussian curvature $K$ of the Riemannian
surface $(\SS ,h)$ as a function of the coordinates $(y,z)$. The masses
are chosen to be $M_1 =M_2 = 1$ for this plot; but the qualitative features
of $K$ are identical for all positive masses. In particular, $K$ is
strictly negative throughout $\SS$, and approaches zero in all three
asymptotic regions, i.e., both
as $\vec{r} \rightarrow \infty$ and as $\vec{r}
\rightarrow \vec{r}_i$, $i=1, \; 2$.

\noindent{\bf Figure 4.} No surprises in Newtonian gravity for the
connection between negative curvature and chaotic geodesic motion:
With completely integrable geodesic flows, 
Gaussian curvature of the metric $g_E$
[Eq.\,(22)] is positive for both the Kepler problem (plot on the left;
strictly positive $K$) and the Newtonian two-center problem (plot on the
right; $K$ positive except in a small neighborhood of the centers).

\noindent{\bf Figure 5.} The Riemannian surface $\SS$ with the metric
corresponding to the null geodesic flow of a single extremal
Reissner-Nordstrom black hole has strictly negative Gaussian curvature,
and the Liapunov exponent of its (unique) closed geodesic is positive.
But there is no trace of chaos here: with angular momentum as the second
integral of motion, the geodesic flow of this surface is a completely
integrable Hamiltonian system.

\noindent{\bf Figure 6.} Construction of the compact set $D \subset \SS$
used in defining the compact future-invariant subset $\Lambda
\subset T_1^\ast \SS$ [see Eq.\,(29)]. The circles $\SC$ and $\SC_i$ are
chosen large enough
so that any geodesic crossing them in the outward direction never
comes back (it escapes to the corresponding asymptotic infinity). The
subset $D$ is the compact connected region bounded
by the three circles.

\noindent{\bf Figure 7.} Closed (or almost closed) orbits in the
geodesic flow on $(\SS ,h)$. The top four plots are drawn with unit
masses $M_1 = M_2 = 1$, and the two plots at the bottom of the figure
are drawn with masses $M_1=2$ and $M_2=1$. As the orbits get more
complicated (and therefore their periods become longer), numerical
instabilities set in as soon as or before the full shape of the orbit
becomes apparent (as happens in the middle two plots). Recall that all
these orbits are unstable because of the negatively curved geometry of
$(\SS ,h)$.

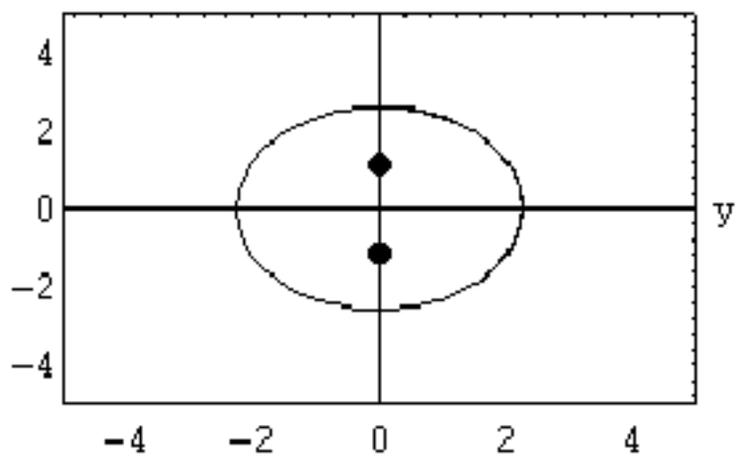
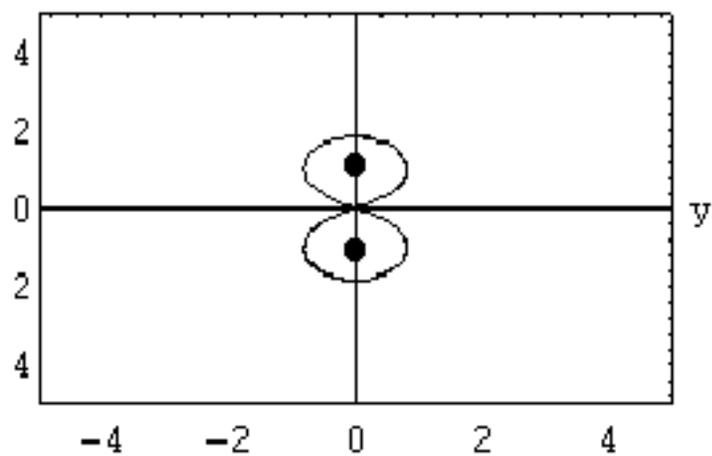
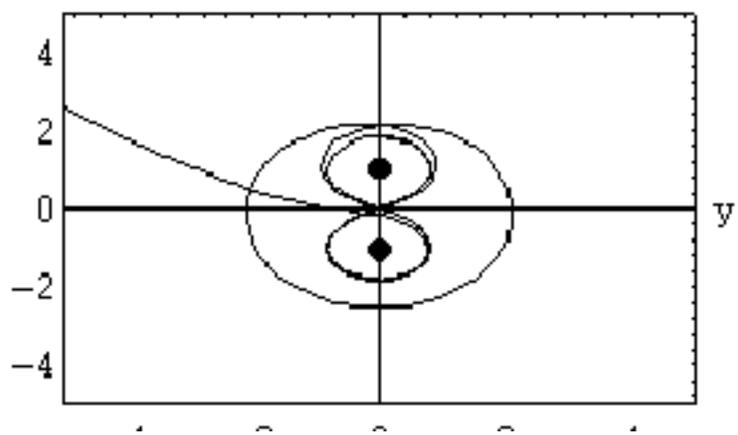
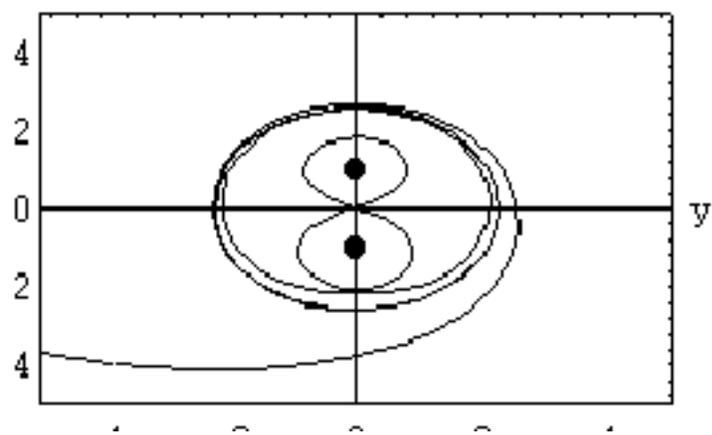
\end{document}